\documentclass{epl}
\newcommand \beq{\begin{eqnarray}}
\newcommand \eeq{\end{eqnarray}}
\newcommand \be{\begin{eqnarray}}
\newcommand \ee{\end{eqnarray}}
\newcommand{\gs}{\stackrel{>}{<}}
\newcommand{\sg}{\stackrel{<}{>}}
\newcommand{\set}[2]{\newcommand{#1}{#2}}
\set{\pa}{\partial \over \partial\, }
\set{\leftvector}{\stackrel{\leftarrow}{\partial }}
\set{\rightvector}{\stackrel{\rightarrow}{\partial }}
\title{Effective mass in quasi two-dimensional systems}
\author{Klaus Morawetz\inst{1,2}}
\institute{
  \inst{1} Institute of Physics, Chemnitz University of Technology, 09107 Chemnitz, Germany\\
  \inst{2} Max-Planck-Institute for the Physics of Complex Systems, 01187 Dresden, Germany
}

\begin{document}

\maketitle

\begin{abstract}
The effective mass of the quasiparticle excitations in quasi two-dimensional systems is calculated
analytically. It is shown that the effective mass increases sharply
when the density approaches the critical one of metal-insulator
transition. This suggests a Mott-type of transition rather than an
Anderson-like transition. The experimental measurements can be reproduced in this way without any additional parameter.
\end{abstract}

The explanation of the metal to insulator transition (MIT) at low
temperatures in quasi two-dimensional systems is still a
strongly debated task. A critical discussion of different approaches
can be found in Ref.~\cite{AKS01}.
The generic feature of MIT transition is the
rapid change from
insulating to conducting behavior when the density is increased
very slightly at low temperatures. This density driven MIT
transitions are usually referred to as Mott transitions. The
characteristic feature of the Mott-Hubbard transition is that the
increasing effective mass is the reason for increasing resistivity
$\rho=m/e^2 n\tau$ while the Anderson scenario would assume a
vanishing relaxation time $n \tau$. It is obvious that this feature characterizes the transition rather than the nature of the insulating state itself, see for details \cite{Ge97}. 

In a recent experiment \protect\cite{SKDK01} it was shown that the
effective mass is increasing sharply when approaching the critical
density. This would underline the Mott picture rather than the
Anderson transition.

Here in this letter we want to substantiate this picture by a
quantitative explanation of the experimental values of the effective
mass. To this end we will introduce a new approximation which is
based on the large mass difference between transport electrons and
scattering impurity of donor ions. Our model consists in electrons
scattering with heavy ions within the quasi two-dimensional gas. In this way we will describe the transition due to Coulomb correlation and not the nature of the insulating state itself. Assuming
the motion restricted to the $x-y$ plane, the
Coulomb potential in this cylindrical Fermi surface is $V_{ab}(q_x,q_y)
= {2 \pi e_a e_b \hbar/\sqrt{q_x^2+q_y^2}}$.

We want to determine the quasiparticle mass which
will be compared to the experimental results.
To this end we will use the standard quasiparticle picture based on
the Green's function method. As generally known, within this approach
the quasiparticle energy and the mass of the model are determined by
the real part of the selfenergy  according to the following formulas.

First, the quasiparticle energy $\hbar \omega=\epsilon_k$
is given as a solution of [$\Sigma={\rm Re} \Sigma^R$]
\be
\omega-{k^2\over 2 m}-\Sigma(k,\omega)=0.
\label{quasi}
\ee

Second, from (\ref{quasi}) the effective mass $m^*$ follows as
\be
{k\over m^*}&=&{\partial \epsilon_k\over \partial k}
={k\over
  m}+{\partial\Sigma\over \partial \hbar \omega}{\partial \epsilon_k\over
  \partial k}+{\partial \Sigma \over \partial k}
={{k\over
  m}+{\partial \Sigma \over \partial k}\over 1-{\partial\Sigma\over
  \partial \hbar \omega}}
\label{mass}
\ee
where the arguments have to be put on-shell $\hbar \omega=\epsilon_k$ after
performing the derivatives.

\begin{figure}
\onefigure[width=9cm]{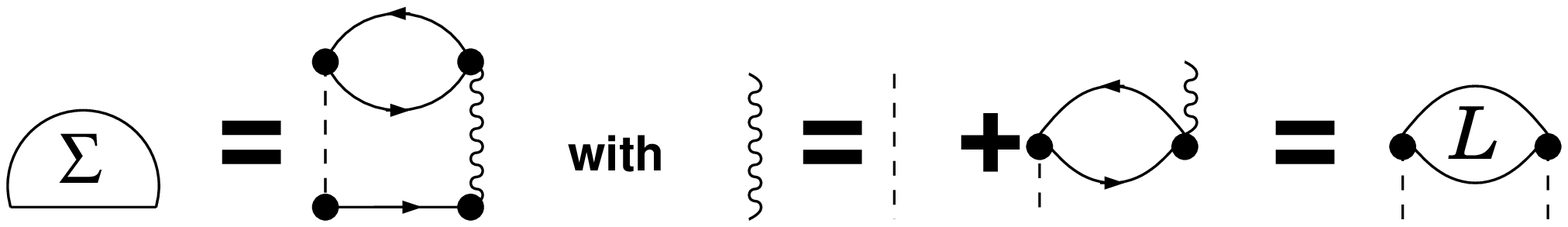}
\caption{\label{vs}The selfenergy
 in terms of
  the screened potential ${\cal V}=V+V \Pi{\cal V}$ (wavy line).}
\end{figure}

In order to proceed we have to know the selfenergy $\Sigma^R$.
For the scattering of electrons on charged ions or holes
it is important to consider the particle-hole fluctuations up to any order.
This is considered in dynamically screened approximation (GW) which expresses
the selfenergy by a sum of all ring diagrams, given in figure~\ref{vs}.
Due to the big mass difference between electrons and ions the vertex corrections are suppressed. The experimental data we want to compare with are characterized by a relatively low density of electrons such that the electron-electron contribution can be neglected.

To avoid technical complications with the Matsubara frequencies and the 
subsequent analytical
continuation to the real time, we will use directly the method of the real 
time nonequilibrium
Green's function even for the equilibrium though its strength lies in the description of nonequilibrium situations.
Within this approach the retarded selfenergy of a species noted by the subscript {\em a} can be written as
\be
\Sigma^R_a(q,\omega)&=&\int{d \omega'\over 2 \pi} {{\cal P}\over
  \omega-\omega'} \Gamma_a(q,\omega')-{i\over 2} \Gamma_a(q,\omega)
\label{sr}
\ee
with the imaginary part
\be
\Gamma_a(q,\omega)=\Sigma^>_a(q,\omega)+\Sigma^<_a(q,\omega).
\label{gamma}
\ee
The correlation parts of the selfenergy, $\Sigma^{\gs}$, in the dynamical
screened approximation read
\be
\Sigma^{\gs}_a(k,\omega)&=&\int\!\! {d q \over (2 \pi \hbar)^2}{d
    \omega'\over 2 \pi} {\cal V}^{\gs}_{a}(q,\omega') G^{\sg}_a(k-q,\omega'\!-\!\omega).
\label{sgtr}
\ee
The screened potential ${\cal V}$ is expressed via the density
fluctuation ${L}$ according to figure~\ref{vs} as
\be
{\cal V}^{\gs}_a(q,\omega)\!&=&\!\sum\limits_bV(q)_{ab}^2 {L}^{\gs}_b(q,\omega)\!=\!\sum\limits_bV(q)_{ab}^2 {\Pi^{\gs}_b(q,\omega)\over |{\cal E}(q,\omega)|^2}
\ee
where ${\cal E}^R(q,\omega)=1-\sum_bV_{bb}(q) \Pi^R_b(q,\omega)$ is the dielectric
function and the polarization or free density fluctuation is given
by
\be
\Pi^{\gs}_a(q,\omega)&=&
\int\! {d p \over (2 \pi \hbar)^2}{d
    \omega'\over 2 \pi} G^{\gs}_a(p,\omega')
  G^{\sg}_a(p\!+\!q,\omega'\!-\!\omega).
\label{pgtr}
\ee

Since we are on the level of the quasiparticle picture,
we can replace in the above formulas
all electron and hole correlation functions
$G^{\gs}$ by their quasiparticle approximation in equilibrium
\be
G^{\gs}_a(k,\omega)&=&
f^{\gs}_a\delta(\epsilon_k - \hbar \omega)
\label{correl}
\ee
where $f^<_a=f_a^{\rm FD}$, $f^>_a=1-f_a^{\rm FD}$ are given by the Fermi function.

As the result we get for the selfenergy (\ref{sgtr}) in the
dynamical screened quasiparticle approximation the formula
\be
&&\Sigma^{\gs}_a(k,\omega)=\sum\limits_b\int {d q \over (2 \pi \hbar)^2} V_{ab}(q)^2
  f^{\gs}_{a,k-q}
\int {d \omega'}
  \delta{(\epsilon_{k-q}-\hbar \omega-\hbar \omega')}{\Pi^{\gs}_b(q,\omega')\over |{\cal E}(q,\omega')|^2}.
\label{stre}
\ee

Since we want to consider here the screening due to the scattering with
heavy impurities (or heavy ionic traps or heavy Fermions)
we can further approximate this selfenergy.
For this purpose we rewrite the free density fluctuation
or polarization function $\Pi^{\gs}$ in a different form.
In terms of the Bose function $g(x)=1/(\exp{(x/T)}-1)$
the polarization function $\Pi^{\gs}$ reads
\be
\Pi^{\gs}_b(q,\omega)=\pm 2 g_b(\pm \hbar \omega){\rm Im}\Pi_b(q,\omega)
\label{10}
\ee
where the imaginary part is
\be
{\rm Im}\Pi_b(q,\omega)&=&
\pi \int {d p \over (2 \pi \hbar)^2}  (f_{b,p}\!-\!f_{b,p+q})
\delta(\epsilon_p\!-\!\epsilon_{p+q}\!-\!\hbar \omega).
\label{imp1}
\ee
A final simplification of the selfenergy can be achieved in the limit of heavy masses of the scattering impurities. We are allowed to neglect quantum
fluctuations of these impurities which are expressed by $g(\pm \hbar \omega)$ in (\ref{10}) and replace $g(\pm
\hbar \omega)\to \pm T/\hbar \omega$.  The selfenergy
(\ref{stre}) can then be written as [$V_{ab}^2=V_{aa}V_{bb}$]
\be
\Sigma^{\gs}_a(k,\omega)&=&2\int {d q \over (2 \pi \hbar)^2} V_{aa}(q)
  f^{\gs}_{a,k-q}
\int {d \omega' \over \omega'}
  \delta{(\epsilon_{k-q}-\hbar \omega-\hbar \omega')}{\rm Im}{1\over
    {\cal E}(q,\omega')}\nonumber\\
&=&-2\pi T \int {d q \over (2 \pi \hbar)^2} V_{aa}(q)
 {\rm Re} \left ( 1-{1\over {\cal E}(q,0)}\right )
f^{\gs}_{a,k-q}
  \delta{(\epsilon_{k-q}-\hbar \omega)}
\label{stre1}
\ee
and the real part of the selfenergy (\ref{sr}) becomes
\be
\Sigma_a(k,\omega)&=&T \!\int\! {d q \over (2 \pi \hbar)^2} V_{aa}(q)
 {\rm Re} \left ( \!1\!-\!{1\over {\cal E}(q,0)}\! \right )  {{\cal P}\over
  \epsilon_{k-q}\!-\!\hbar \omega}.
\label{13}
\ee

\begin{figure}
\twofigures[width=7cm]{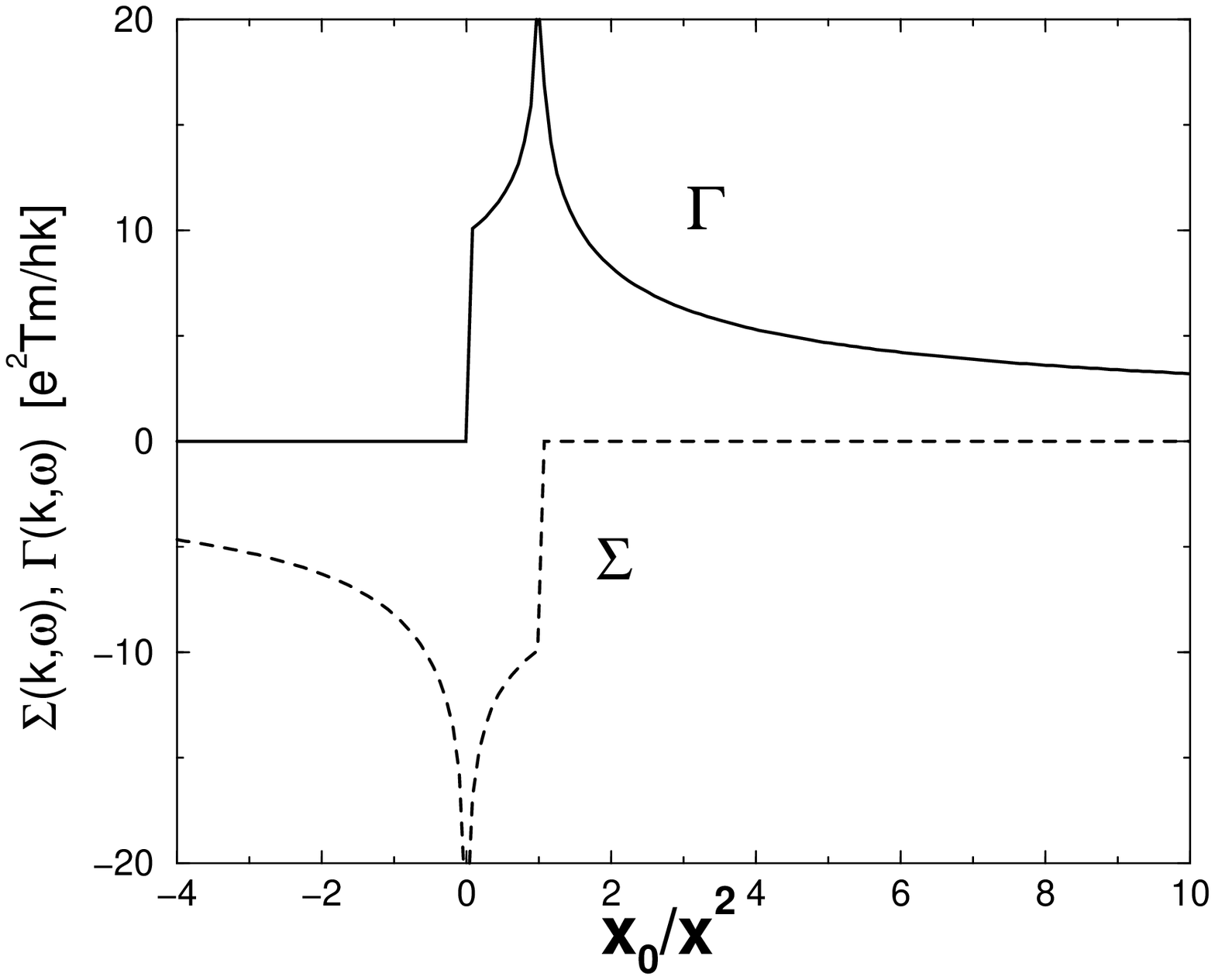}{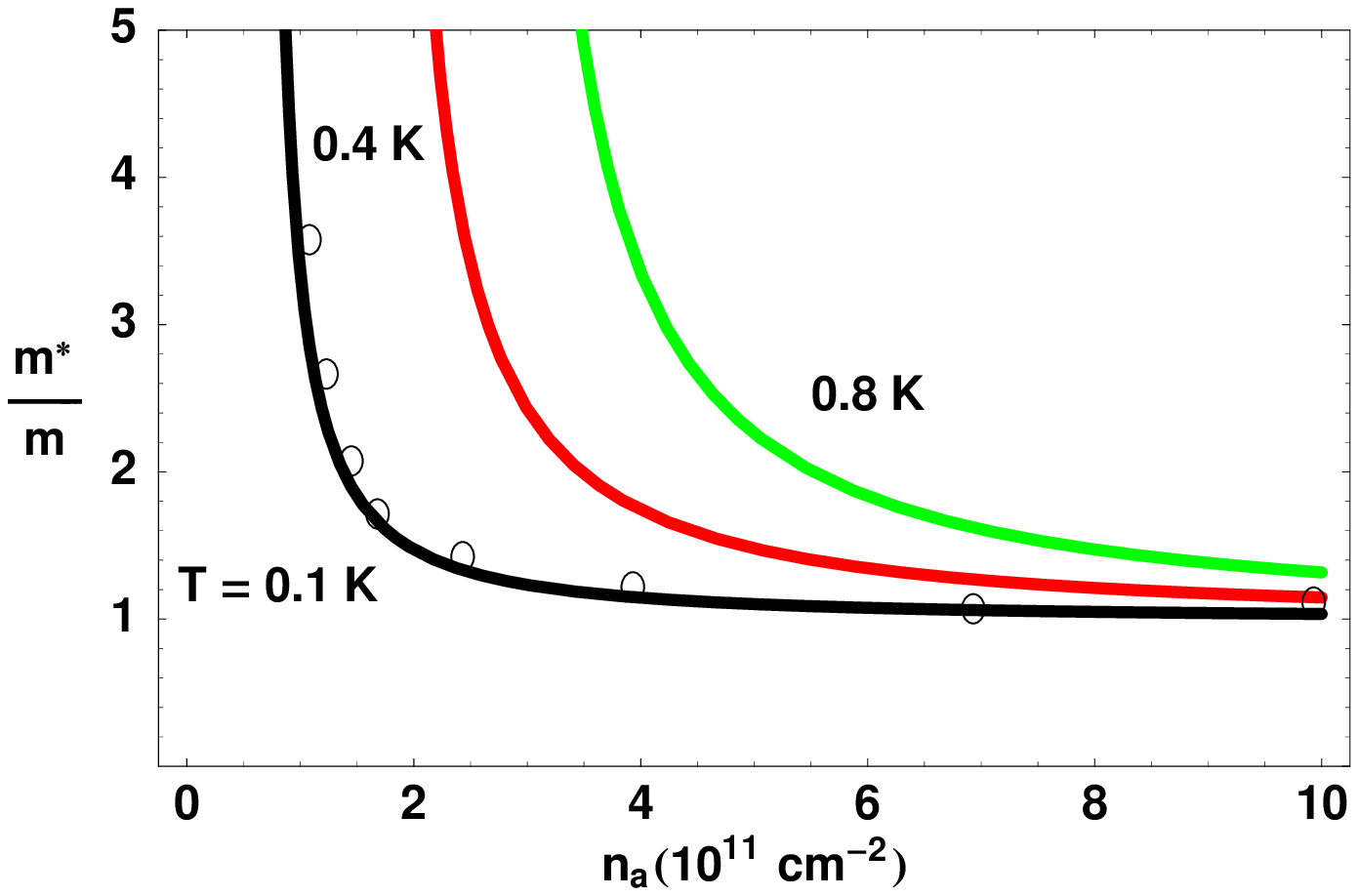}
\caption{\label{massp}The real and imaginary part of the first temperature correction of the selfenergy (\protect\ref{g}) and (\protect\ref{s}) versus scaled frequency $x_0/x^2=2 m \hbar \omega/k^2$.}
\caption{\label{m}The effective mass for temperatures
  $0.1,0.4,0.8$K from left to right according to (\protect\ref{ms}) versus density. The
  open circles are experimental values \protect\cite{SKDK01} at $T=0.1$K.}
\end{figure}

In order to evaluate the effective mass according to (\ref{mass}), it remains
to get an expression for the dielectric function.
To this end we use the zero temperature expansion of the dielectric function in
quasi two-dimensions \cite{HSW71,Mc02} since the leading temperature dependence is already in front of the integral. This means we consider the electrons as degenerated but the impurity particle-hole fluctuation as classical here.
Changing the integration variables $z=q/2 p_{f}$, we then obtain for the real part (\ref{13}) of
the dynamical screened selfenergy [$x=k/2 p_{f}$, $x_0=\hbar \omega/4 \epsilon_{f}$]
\be\label{dynfluc}
\Sigma_a(k,\omega)\!&=&\!-{e_a^2 T m_a\over 2 \hbar p_{f} x} \!\int\limits_0^\infty\! {d z \over {x z\over \kappa_a}
\!+\! 1} 
\int\limits_0^{2\pi}\! {d \phi\over {x_0\over x^2}\!-\!1\!+\!2 z
  \cos{\phi}\!-\!z^2}
\approx
-{e_a^2 T m_a\over 2 \hbar p_{f} x} \!\int\limits_0^\infty\! d z  
\int\limits_0^{2\pi}\! {d \phi\over {x_0\over x^2}\!-\!1\!+\!2 z
  \cos{\phi}\!-\!z^2}
\nonumber\\&=&
{2 \pi e_a^2 T m_a\over \hbar p_{f}} 
\left \{ 
\begin{array}{ll}
\frac{2}{ x+\sqrt{x_0}}{\cal K}\left [ \left
    ({\sqrt{x_0}-x\over \sqrt{x_0}+x}\right )^2\right]
&
0<x_0<x^2
\\
{\pi \over 4 x}
&
x_0=x^2
\\
0
&
x_0>x^2
\\
\frac{1}{\sqrt{x^2-x_0}}{\cal K}\left [ {x^2\over x^2-x_0}\right]
&
x_0<0
\end{array}
\right .
\label{s}
\ee
with ${\cal K}$ the complete elliptic integral of first kind and
$\kappa_a={\hbar \chi_a / 2p_{f}}$. The inverse screening length is given by $\chi_a=2 e_a^2 m_a/\hbar^2$.
The approximate sign concerns the limit of large $\kappa_a$ which is
justified for the parameters used here since for typical densities of $7\times 10^{10}$cm$^{-2}$
we have
\be
\kappa_a={284.9\over \sqrt{n_a/7\times 10^{10}{\rm cm}^{-2}}}
{m^*\over m}.
\ee
In the same way we can evaluate the imaginary part of the selfenergy (\ref{gamma}) with the result
\be
\Gamma_a(k,\omega)={2 \pi^2 e_a^2 T m_a\over \hbar p_{f}}
\frac{1}{x+\sqrt{x_0}}{\cal K}\left [ {4 \sqrt{x_0}\over (x+\sqrt{x_0})^2}\right].
\label{g}
\ee
The linear temperature dependence appears here from the neglect of the 
quantum fluctuations in the ionic particle-hole fluctuation and is due to screening and should not be confused with the standard non-Fermi liquid behavior in the literature. The latter one is more visible in the divergency at the Fermi energy $x_0=x^2$ in figure~\ref{massp}. Later we will give also the statically screened result which can be considered as a Born approximation of an electron scattering with a neutral impurity in the form of a Debye potential. This selfenergy has a finite zero temperature limit but the imaginary part does not vanish at the Fermi energy though it is not diverging. For a Fermi liquid we would expect a vanishing imaginary part of the selfenergy at the Fermi energy. 

The real part $\Sigma$ is related to this imaginary $\Gamma$ by the Hilbert transform (\ref{sr}). The more astonishing is the fact that we find here an additional 
relation
\be
\Sigma(k,\omega)=-\Gamma(k,{k^2\over 2 m \hbar}-\omega)
\ee
which is only valid for this specific type of selfenergy besides the Kramers-Kronig relation (\ref{sr}). From the  plot in figure~\ref{massp} we see that the excitation are only possible for positive frequencies since we have calculated the first temperature correction. One can see that this first temperature correction has a highly nontrivial frequency behavior far from being Fermi-liquid like.

We can now evaluate the effective mass
and compare its dependence on the density with the experimental facts in figure~\ref{m}.
Since the real part of the selfenergy (\ref{dynfluc}) has the dependence on momentum and frequency as $\Sigma \propto 1/k
F [2 m \omega/k^2]$ we see that (\ref{mass}) takes exactly the form
\be
{m\over m^*}={1-{1\over 4
    \epsilon_{f}}\Sigma_a(p_{f},\epsilon_{f})\over
1+{1\over 4
    \epsilon_{f}}\Sigma_a(p_{f},\epsilon_{f})}.
\label{ms}
\ee
The needed on-shell value of the self energy (\ref{dynfluc}) simplifies
\be
{\Sigma_a(p_{f},{p_{f}^2 \over 2 m_a})\over \epsilon_{f}}&=&\pi^2
{T\over \epsilon_f}
\kappa_a
\label{ss1}
\ee
with
$\kappa_a={\hbar \chi_a\over 2p_{f}}={e_a^2 m_a\over \hbar
  p_{f}}$.
According to the formulae (\ref{ms}) we obtain a sharp increase of the
effective mass for lower densities (figure~\ref{m}) still above the critical 
one. This is in very good agreement with the
measurements \cite{SKDK01}. 

In figure~\ref{m} we compare the formula (\ref{ms}) with (\ref{ss1}) with the experimental values of \cite{SKDK01}. The agreement is astonishing precise considering the fact that the only parameter entering the formulas are the experimental density and temperature.

The temperature dependence of the effective mass turns out to be linear over a wide range since 
from (\ref{ms}) and (\ref{ss1}) we obtain for small corrections $\Sigma$
\be
{1\over m*}&\approx&{1\over m} \left (1-{1\over 2 \epsilon_f}\Sigma_a(k,{k^2\over 2
    m})\right )_{k=p_f}
={1\over m} \left (1- \pi^2 {T\over \epsilon_f} \kappa_a\right )
\label{effmass}
\ee
while for large $\Sigma$ the equation (\ref{ms}) has to be used.
This implies a nearly linear temperature dependence for the conductivity, $\sigma=n e^2 \tau /m^*$, if the relaxation time behaves accordingly. This has been shown indeed in \cite{Mc02}.
The linear temperature dependence
has been repeatedly reported in the literature both from experimental and
theoretical point of view.
Numerical calculations of Coulomb scattering rates 
on impurities predict a linear temperature dependence of the 
mobility in silicon inversion layers \protect\cite{CW80,SH99}.
This was attributed to the collisional level broadening in the 
screening function. Related results have been obtained in 
Ref.~\cite{S86} where a substantial suppression of the temperature 
dependence in the screening function was found. An analytical 
investigation of screening in quasi two-dimensional systems was 
given in Ref.~\cite{GD86} where a linear temperature term to the 
conductivity was found which was confirmed in \cite{Mc02}.

When the density decrease towards the critical value, the electrons become trapped by the charged impurities forming neutral impurities. This is the mechanism of Mott transition. A corresponding mass action law will then determine the neutral impurity concentration $n_i$. Above the critical density these neutrals are negligible. We suggest here that the experimental values in figure \ref{m} are well above this value. 

For the reason of completeness we give now the contribution to the effective mass if there is a scattering with the neutral impurities. We can use a Debye form of the scattering potential which results formally from the static approximation in (\ref{stre})
\be
V_i(q)={V(q) \over
  |{\cal E}(q,0)|}={2 \pi g_{ab}\hbar \over q+{\hbar \over r_0}}
\label{vs1}
\ee
if we introduce the scattering strength $g_{ab}=e_a e_b$ and the range of interaction $r_0=1/\kappa$. In the second Born
approximation \cite{F94} the relation between the scattering strength and the scattering length reads
\be
a_0&\approx&-{2 m\over \hbar^2} {g_{ab} r_0^2} \left ( 1+ {m\over
    \hbar^2} {g_{ab} r_0}\right ).
\ee
From (\ref{stre}) we obtain now 
\be
\Sigma_i^{\gs}(k,\omega)&=&2 \pi n_i\int{d q \over (2 \pi \hbar)^2} {V_i(q)^2 }
f^{\gs}_{k-q}\delta{(
  \hbar \omega-\epsilon_{k-q})}
\ee
and instead of (\ref{dynfluc}) the real part of the self energy reads
\be\label{statfluc}
\Sigma_i(k,\omega)&=&{g_{ab}^2 n_i \pi \over 4 \epsilon_{f} x^2}  
 \int\limits_{2
  {\sqrt{x_0}\over x}}^\infty
{ds\over ({\kappa_i\over x^2}+\sqrt{s+{x_0\over x^2}+1})^2}{1\over\sqrt{s^2-{4 x_0\over x^2}}}.
\ee
The on-shell value $x_0=x^2$ can be performed analytically to
\be
\Sigma_i(k,{k^2 \over 2 m})&=&-{g_{ab}^2  m_a x^2\over  2\hbar^2}
{n_i\over n_a} \left ( {1\over 4 x^4-\kappa_i^2}-{\kappa_i {\rm arccos}{\kappa_i\over 2 x^2}\over
    (4 x^4-\kappa_i^2)^{3/2}}  \right
)\nonumber\\
&=&-{g_{ab}^2  m_a x^2 \over  2\hbar^2}
{n_i\over n_a}\left ( 
{\ln{{\kappa_i\over x^2}}-1\over\kappa_i^2} \right )+o({1\over \kappa_i^3})
\label{sss}
\ee
with $\kappa_i=\hbar/2 r_0 p_{f}$. The value at the Fermi energy is
\be
{\Sigma_i(p_{f},{p_{f}^2 \over 2 m_a})\over \epsilon_{f}}=-{4} {n_i\over n_a} \left ({m g_{ab} r_0\over \hbar^2}\right )^2 (\ln{\kappa_i}-1).
\label{sss1}
\ee
Together with the electron -- charged impurities contribution (\ref{effmass}) we obtain finally
\be
{m\over m*}&=&\left (1- \pi^2 {T\over \epsilon_f} \kappa_a+{4} {n_i\over n_a} \left ({m g_{ab} r_0\over \hbar^2}\right )^2 (\ln{\kappa_i}-1)\right ).
\label{effmass1}
\ee

When the neutral impurities become significant near the Mott-transition we have to use a mass action law to determine $n_i$. This can be found from a simplified semiconductor model. Assuming only the scattering with donor levels, 
the total donor concentration consists of neutral and charged 
impurities, $n=n_i+n_i^+$. The neutral donors correspond to 
electrons trapped at the donor level $E_D$. With the Fermi
energy $E_F$, from the thermal population follows \cite{AM76}
\be
n_i={n\over \frac 1 2 {\rm e}^{E_D-E_F\over T}+1}
\label{d1}
\ee
where the factor $1/2$ comes from the two possible states
at each impurity site.
The electron density in the band is given by the effective 
conduction band energy $E_c(n,T)$, \cite{IL96}
\be
n_i^+\approx n_a  ={m_a T\over \pi \hbar^2} 
{\rm e}^{E_F-E_c(n_a,T)\over T}.
\label{d2}
\ee 
where $E_c(n_a,T)$ describes effective conduction band level which
becomes density and temperature dependent by the correlation effects of the 
electrons. 
Eliminating the Fermi energy in (\ref{d1}) and (\ref{d2}) 
one obtains the mass action law \cite{IL96}
\be
{n_i\over n_a}={{2 \pi \hbar^2 n_a\over m_a T}} 
{\rm e}^{E_c(n_a,T)-E_D \over T}.
\label{mott}
\ee
The density dependence of the selfenergy $\Sigma$ which determines the
effective conduction band $E_c$ leads to 
a nonlinear density dependence of this ratio. In principle also the
donor level $E_D$ is density dependent. But, we can safely condense
both effects into an effective density dependent binding energy 
$E_b(n_a,T)=E_c(n_a,T)-E_D$.
With increasing density of the electrons more collisions with donors
happens and the formation of bound states
is favored until a critical density is reached where 
pressure ionization happens.
At this critical density the trapped states are resolved 
called here the Mott-transition. The particle 
distance becomes smaller than the Bohr radius. 

As conclusion, the found effective mass here which can describe the experimental values seems to support the 
Mott-transition picture rather than the Anderson scheme, the same conclusion is obtained from the calculation of the conductivity \cite{Mc02} where the reader is kindly referred to for a more detailed discussion.
It should be pointed out however, that the nature of the insulating state itself is not clarified by the above consideration though we can describe the sharp increase of effective mass by Coulomb correlations. The suggested trapping mechanism of electrons on charged impurities \cite{Mc02} is not the only possible one. An alternative is the formation of three-particle clusters \cite{NM01} which is more favourable in high magnetic fields. The strong magnetic field dependence seems to underline such idea. Further insight will be gained from the calculation of magnetic field dependence than the increase of effective mass which is shown merely due to Coulomb correlations.

The enlightening discussions with Peter Fulde, Pavel Lipavsk\'y, Enver Nakhmedov, Debanand Sa, Michael Schreiber and 
Va\v clav \v Spi\v cka are gratefully acknowledged.


\begin{thebibliography}{10}

\bibitem{AKS01}
E. Abrahams, S.~V. Kravchenko, and M.~P. Sarachik, Rev. Mod. Phys. {\bf 73},
  251  (2001).

\bibitem{Ge97}
F. Gebhard, {\em The Mott Metal-Insulator Transition} (Springer, Berlin, 1997).

\bibitem{SKDK01}
A.~A. Shashkin, S.~V. Kravchenko, V.~T. Dolgopolov, and T.~M. Klapwijk, Phys.
  Rev. B {\bf 66},  073303  (2002).

\bibitem{HSW71}
C. Hodges, H. Smith, and J.~W. Wilkins, Phys. Rev. B {\bf 4},  302  (1971).

\bibitem{Mc02}
K. Morawetz, Phys. Rev. B {\bf 67},  115125  (2003).

\bibitem{CW80}
F. Stern, Phys. Rev. Lett. {\bf 44},  1469  (1980).

\bibitem{SH99}
S. DasSarma and E.~H. Hwang, Phys. Rev. Lett. {\bf 83},  164  (1999).

\bibitem{S86}
S. DasSarma, Phys. Rev. B {\bf 33},  5401  (1986).

\bibitem{GD86}
A. Gold and V.~T. Dolgopolov, Phys. Rev. B {\bf 33},  1076  (1986).

\bibitem{F94}
S. Fl{\"u}gge, {\em Practical Quantum Mechanics} (Springer, Berlin, 1994).

\bibitem{AM76}
N.~W. Ashcroft and N.~D. Mermin, {\em Solid State Physics} (Saunders College,
  Philadelphia, 1976).

\bibitem{IL96}
H. Ibach and H. L{\"u}th, {\em Solid state physics: An introduction to
  principles of materials science} (Springer Verlag, Berlin, 1996).

\bibitem{NM01}
E. Nachmedov and K. Morawetz, Phys. Rev. B {\bf 66},  195333  (2002).

\end{thebibliography}

\end{document}